\documentclass[11pt,a4paper]{article}
\usepackage[margin=27mm]{geometry}
\usepackage{amsmath,amssymb,amsthm,graphicx,url,placeins,capt-of}
\usepackage{comment}

\numberwithin{equation}{section}
\newtheorem{theorem}{Theorem}[section]

\usepackage{algorithm,algpseudocode}
\usepackage{needspace}
\usepackage[hidelinks]{hyperref}
\title{Exact Inversion of Relaxation from Rheometer\\Measurement Data\\[1ex]{\large\itshape Dedicated to the late Professor Anatoly Grigorievich Yagola}}
\author{Yu Jiang\thanks{Corresponding author. School of Mathematics, Shanghai University of Finance and Economics, 777 Guoding Road, Shanghai 200433, China. E-mail: \url{jiang.yu@mail.shufe.edu.cn}.}\quad Takayuki Kurokawa\thanks{Graduate School of Advanced Life Sciences, Hokkaido University, Sapporo 060-0810, Japan. E-mail: \url{kurokawa@sci.hokudai.ac.jp}.}\quad Ching-Lung Lin\thanks{Department of Mathematics, National Cheng-Kung University, Tainan 701, Taiwan. E-mail: \url{cllin2@mail.ncku.edu.tw}.}\quad Gen Nakamura\thanks{Department of Mathematics, Hokkaido University, Sapporo 060-0808, Japan, and Research Center of Mathematics for Social Creativity, Research Institute for Electronic Science, Hokkaido University, Sapporo 060-0812, Japan. E-mail: \url{gnaka@math.sci.hokudai.ac.jp}.}}
\date{}
\begin{document}
\maketitle
\pagestyle{myheadings}
\markright{Exact Inversion of Relaxation from Rheometer Data}
\begin{abstract}
A rheometer is commonly used to measure the input-output relation for analyzing the relaxation of stress for rheological materials. Here, the input is a time-harmonic torque and the output is the angular displacement. This paper aims to propose an exact inversion scheme of the relaxation function from two input-output relations only having different moments of inertia.
The relaxation function is represented by a Prony's series of known number of terms, with a positive constant term and terms with distinct relaxation times. The augmented system relates the material parameters to its characteristic roots. These roots are the poles of the transfer function. Here the transfer function is the Laplace transform of the output divided by that of the input, and it is a rational function in the Laplace variable. The poles of the transfer function are recovered using Pad\'e's approximation. By measuring two input-output relations only having different moments of inertia, we can rigorously recover the relaxation function. Thus, we note a clear link between the characteristic roots and the poles. Numerical experiments reconstruct the relaxation function from synthetic angular displacements corresponding to two given moments of inertia based on our inversion method. The results show the effect of measurement noise on the recovered characteristic roots and relaxation function.
\end{abstract}
\noindent\textit{2010 Mathematics Subject Classification:} 45K05, 45D05.

\noindent\textit{Keywords:} inverse problem, relaxation function, viscoelasticity, rotational rheometer, Prony series, Pad\'e's approximation.


\section{Introduction}
The relaxation function describes how a viscoelastic material retains and dissipates stress over time. Identifying this function provides a link between measured mechanical responses and constitutive models \cite{hamley}. Finite sums of positive decaying exponentials and a positive constant offer a practical representation in terms of relaxation times and moduli. Such representations in terms of Prony series have been used both to approximate relaxation functions and to construct models from creep data \cite{mauro,luo}. The relaxation times $\tau_i$, the moduli $G_i$ and the equilibrium modulus $G_e$ are referred to as the relaxation parameters. Recovering the relaxation function is a distinct objective in rheology.

Since we have already given in Abstract a brief explanation of our method to recover a relaxation function using a rheometer, let us first give a brief overview of several other methods to recover a relaxation function or extract some of its property. In rheology, material responses may be measured in the time domain or through oscillatory tests. Optimally windowed chirps provide an example of rapid acquisition of dynamic response over a range of frequencies \cite{geri}. Honerkamp and Weese \cite{honerkamp} developed nonlinear regularization to analyze the different contributions of short and long relaxation times. Parameter estimation for exponential sums has also been studied through approximate Prony's methods \cite{potts}. These studies motivate careful attention to both the available data and the sensitivity of the recovered parameters.

We want to recover a relaxation function modeled as a Prony's series with a positive constant term (\cite{luo}) from the measured data using a stress controlled rheometer. More precisely, we want to propose an exact inversion scheme to recover Prony's series from two input-output relations only having different moments
of inertia. The mathematical model of measurement is specified below.


We consider small deformations in the linear viscoelastic regime, with fixed geometry and no slip. Sample inertia, edge effects and instrument compliance are neglected. For an initially relaxed sample, the inertial stress balance and Boltzmann superposition principle in Eqs. (2) and (4) of Ogunkeye et al.~\cite{ogunkeye} give the following equation after conversion from stress and strain to torque and angular displacement.
\begin{equation}
I\ddot\theta(t)+\int_0^tG(t-s)\dot\theta(s)\,ds=T(t)\quad\text{for }t\ge0,\label{eq:motion}
\end{equation}
Here $\theta$ is the angular displacement, $T$ is the applied torque, and $I>0$ is the total rotating inertia of the instrument and fixture. The rotational relaxation kernel is $G=b\,G_{\mathrm{mat}}$, where $G_{\mathrm{mat}}$ is the material shear relaxation modulus. The known geometry factor is $b=2\pi R^3/(3\tan\alpha)$ for a cone of radius $R$ and small angle $\alpha$, and $b=\pi R^4/(2h)$ for ideal parallel plates of radius $R$ and gap $h$ \cite{ewoldt}. The material modulus is $G/b$. The same geometry and material state are retained when the inertia is changed. We note that there are different rheometer configurations and corresponding measurement models and control modes, as discussed in \cite{training,lauger2010,lauger2016,rheocompass} and the references therein.

We assume a Prony's series representation of the relaxation function \cite{luo,mauro,kammler},
\begin{equation}G(t)=G_e+\sum_{i=1}^NG_i\exp\left(-\frac{t}{\tau_i}\right),\label{eq:prony}\end{equation}
where each $G_i,\tau_i$ and $G_e$ are positive constants. We assume that relaxation times $\{\tau_i\}_{i=1}^N$'s are distinct. In the extended Maxwell model with $N$ Maxwell elements (\cite{luo}), $G_i$ and $\eta_i$ are, respectively, the spring and viscosity coefficients of the $i$-th element after multiplication by $b$. Likewise, $G_e$ is the equilibrium rotational stiffness. The material moduli are $G_i/b$ and $G_e/b$. Further, each $\tau_i$ is given as $\tau_i=\eta_i/G_i$. As for $T(t)$, we assume that it is given as
\begin{equation}T(t)=T_0\sin(\omega_0t+\phi).\label{eq:torque}\end{equation}
with angular frequency $\omega_0>0$, torque amplitude $T_0>0$, initial phase $\phi\ge0$. Finally, we assume that the initial condition for $\theta(t)$ is given as
\begin{equation}\theta(0)=0,\ \dot\theta(0)=0.\label{eq:initial}\end{equation}
We remark here that there may be an additional term $Q\,\theta(t)$ in the lefthand side of \eqref{eq:motion} with a known constant $Q>0$ describing a structural rotational elasticity of the rheometer. However, since it can be absorbed into $G_e$, we have omitted this term.


In \cite{chen1, chen2}, the authors built a spectral approach using a clustered eigenvalue for \eqref{eq:motion} without $T(t)$ to exactly recover the relaxation function. For each $I$, the characteristic roots mentioned in Abstract are clustered eigenvalues.
As we revealed in Abstract, there is a close relation between the spectral approach and the input-output relation approach. This paper applies the spectral approach via this relation.

The paper is organized as follows. In Section 2, by introducing internal state variables $q_i(t)$'s, we transform \eqref{eq:motion} into a first-order system called the augmented system (\cite{chen1})
\begin{equation}\dot{\mathbf{x}}(t)={\mathbf{A}\mathbf{x}}(t)+{\mathbf{B}}T(t)\label{eq:augmented}\end{equation}
where $\mathbf{A}$ is a $(N+2)\times (N+2)$ matrix and $\mathbf{B}$ is a vector.
Then, we derive the characteristic polynomial of the augmented system in two ways: one by deriving the equation for $\dot\theta$ and the other by directly computing the characteristic equation of \eqref{eq:augmented}. Also, we analyze the structure of its roots. Section 3 is devoted to giving an exact inversion scheme of the relaxation function assuming that both moments of inertia are known. More precisely, by knowing two input-output relations associated with two different moments of inertia, we give an exact inversion scheme to recover the relaxation function. Section 4 presents the reconstruction algorithm and tests both steps using synthetic angular displacements with and without noise. Section 5 summarizes the main findings and discusses future work.


\section{Augmented system and its characteristic roots}
In this section, by introducing internal state variable functions $q_i(t)$'s, we transform \eqref{eq:motion} into a first-order system of ordinary differential equations \eqref{eq:augmented}.
\begin{equation}I\ddot\theta(t)+G_e\theta(t)+\sum_{i=1}^Nq_i(t)=T(t),\label{eq:old2-1}\end{equation}
where
$q_i(t):=G_i\int_0^t e^{-(t-u)/\tau_i}\dot\theta(u)\,du$. Differentiation each $q_i(t)$ gives
\begin{equation}\dot q_i(t)+\frac1{\tau_i}q_i(t)=G_i\dot\theta(t)\quad\text{for }i=1,2,\ldots,N\label{eq:old2-2}\end{equation}

Now, let
\begin{equation}\mathbf{x}(t)=(x_1(t),x_2(t),x_3(t),\dots,x_{N+2}(t))^{\mathfrak{t}}
=(\theta(t),\dot\theta(t), q_1(t),\dots,q_N(t))^{\mathfrak{t}},
\end{equation}
where the notation $\mathfrak{t}$ denotes the transpose of vectors. Then,
the explicit forms of $\mathbf{A}$ and $\mathbf{B}$ in \eqref{eq:augmented} are given as
\begin{equation}\label{eq:system-matrix}\mathbf{A}=\begin{bmatrix}0&1&0&\cdots&0\\-G_e/I&0&-1/I&\cdots&-1/I\\0&G_1&-1/\tau_1&\cdots&0\\\vdots&\vdots&\vdots&\ddots&\vdots\\0&G_N&0&\cdots&-1/\tau_N\end{bmatrix}
\end{equation}
and
$\mathbf{B}=(0, 1/I,0,\dots,0)^\mathfrak{t}$.

To determine the eigenvalues of $\mathbf{A}$, let $\mathbf{x}\ne{\mathbf{0}}$ satisfy
$(\lambda{\mathbf{Id}}-{\mathbf{A}}){\mathbf{x}}={\mathbf{0}}$. Its component equations are
\[\begin{cases}\lambda x_1-x_2=0,\\\dfrac{G_e}{I}x_1+\lambda x_2+\displaystyle\sum_{i=1}^N\dfrac1I x_{i+2}=0,\\-G_ix_2+\left(\lambda+\dfrac1{\tau_i}\right)x_{i+2}=0\quad\text{for }i=1,2,\ldots,N.\end{cases}\]
Before dividing by $\lambda$ or $\lambda+1/\tau_i$,
we exclude their zero values directly from these component equations.

If $\lambda=0$, the first equation gives $x_2=0$. The internal equations then give $x_{i+2}=0$ for every $i\,{(1\le i\le N)}$. The second equation gives $G_ex_1=0$. Hence $x_1=0$ because $G_e>0$. This contradicts $\mathbf{x}\ne{\mathbf{0}}$.
If $\lambda=-1/\tau_k$, the $k$-th internal equation gives
$G_kx_2=0$, hence $x_2=0$. Since $\lambda\ne0$, the first equation
gives $x_1=0$. Distinct relaxation times imply
$\lambda+1/\tau_i\ne0$ for $i\ne k$, so $x_{i+2}=0$ for these
indices. The second equation now gives $x_{k+2}=0$, again a contradiction.
Thus neither zero nor any $-1/\tau_i$ is an eigenvalue of $\mathbf{A}$.

For every remaining candidate $\lambda$, the first and internal equations
give $x_1=x_2/\lambda$ and
$x_{i+2}=G_i\tau_i x_2/(1+\tau_i\lambda)$. Substitution into the
second equation yields
\[\frac{G_e}{I}\left(\frac1\lambda\right)x_2+\lambda x_2+\sum_{i=1}^N\frac1I\left(\frac{G_i\tau_i}{\tau_i\lambda+1}\right)x_2=0.\]
The preceding expressions show that $x_2=0$ would
force $\mathbf{x}={\mathbf{0}}$. Hence $x_2\ne0$, and division by $x_2$ gives
\begin{equation}\lambda+\frac{G_e}{I\lambda}+\sum_{i=1}^N\frac{G_i\tau_i}{I(\tau_i\lambda+1)}=0.\label{eq:old2-5}\end{equation}
Multiplying \eqref{eq:old2-5} by $I\lambda\,\Pi_{j=1}^N(\tau_j\lambda+1)$, we obtain the characteristic equation, with the subscript $I$ denoting the moment of inertia,
\begin{equation}{P_I}(\lambda):=(I\lambda^2+G_e)\prod_{j=1}^N(\tau_j\lambda+1)+\sum_{i=1}^N\left[G_i\tau_i\lambda\prod_{j=1,j\ne i}^N(\tau_j\lambda+1)\right]=0,\label{eq:poly}\end{equation}
and its roots are called characteristic roots.
Using $\tau_j\lambda+1=\tau_j(\lambda+1/\tau_j)$, equation \eqref{eq:poly} gives
\[{P_I}(\lambda)=\left(\prod_{j=1}^N\tau_j\right)
\left[(I\lambda^2+G_e)\prod_{j=1}^N(\lambda+1/\tau_j)
+\sum_{i=1}^NG_i\lambda\prod_{j\ne i}(\lambda+1/\tau_j)\right].\]

Next, we will show that $P_I(\lambda)
$ can be derived by computing $\det(\lambda\mathbf{Id}-\mathbf{A})$. For that, separate the first two variables from the $N$ internal variables and write
\[\lambda{\mathbf{Id}}-{\mathbf{A}}=\begin{pmatrix}{\mathbf{B}}_\lambda&{\mathbf{C}}\\{\mathbf{E}}&{\mathbf{D}}_\lambda\end{pmatrix},\qquad
{\mathbf{B}}_\lambda=\begin{pmatrix}\lambda&-1\\G_e/I&\lambda\end{pmatrix},\qquad
{\mathbf{D}}_\lambda=\operatorname{diag}(\lambda+1/\tau_1,\ldots,\lambda+1/\tau_N).\]
The off-diagonal blocks are the following $2\times N$ matrix $\mathbf{C}$ and $N\times 2$ matrix $\mathbf{E}$
\[{\mathbf{C}}=\begin{pmatrix}0&\cdots&0\\1/I&\cdots&1/I\end{pmatrix},\qquad
{\mathbf{E}}=\begin{pmatrix}0&-G_1\\\vdots&\vdots\\0&-G_N\end{pmatrix}.\]
For $\lambda\ne-1/\tau_i$, the diagonal block $\mathbf{D}_\lambda$ is invertible. Block elimination gives
\[\begin{pmatrix}{\mathbf{Id}}_2&-{\mathbf{C}\mathbf{D}}_\lambda^{-1}\\{\mathbf{0}}&{\mathbf{Id}}_N\end{pmatrix}
\begin{pmatrix}{\mathbf{B}}_\lambda&{\mathbf{C}}\\{\mathbf{E}}&{\mathbf{D}}_\lambda\end{pmatrix}
=\begin{pmatrix}{\mathbf{B}}_\lambda-{\mathbf{C}\mathbf{D}}_\lambda^{-1}{\mathbf{E}}&{\mathbf{0}}\\{\mathbf{E}}&{\mathbf{D}}_\lambda\end{pmatrix}.\]
The leftmost matrix has determinant one. Taking determinants therefore yields
\[\det(\lambda{\mathbf{Id}}-{\mathbf{A}})=\det {\mathbf{D}}_\lambda\,\det({\mathbf{B}}_\lambda-{\mathbf{C}\mathbf{D}}_\lambda^{-1}{\mathbf{E}}).\]
Here the remaining $2\times2$ block is
\[{\mathbf{B}}_\lambda-{\mathbf{C}\mathbf{D}}_\lambda^{-1}{\mathbf{E}}=
\begin{pmatrix}\lambda&-1\\G_e/I&\displaystyle\lambda+\frac1I\sum_{i=1}^N\frac{G_i}{\lambda+1/\tau_i}\end{pmatrix}.\]
Computing its determinant gives
\[\det(\lambda{\mathbf{Id}}-{\mathbf{A}})=\prod_{j=1}^N(\lambda+1/\tau_j)
\left[\lambda^2+\frac{G_e}{I}+\frac1I\sum_{i=1}^N\frac{G_i\lambda}{\lambda+1/\tau_i}\right].\]
Comparison yields
\begin{equation}\det(\lambda{\mathbf{Id}}-{\mathbf{A}})=\frac{{P_I}(\lambda)}{I\prod_{j=1}^N\tau_j}.\label{eq:det-poly}\end{equation}
Both sides are polynomials, so this identity also holds at $\lambda=-1/\tau_i$.
Consequently, $P_I$ is the unnormalized characteristic polynomial, with
the same roots and multiplicities as $\det(\lambda{\mathbf{Id}}-{\mathbf{A}})$.
As a direct check of its exceptional values, its explicit definition gives
\[{P_I}(0)=G_e>0,\qquad
{P_I}(-1/\tau_k)=-G_k\prod_{j\ne k}(1-\tau_j/\tau_k)\ne0.\]

Concerning the properties of characteristic roots, we have the following.

\begin{theorem}Let
\begin{equation}0<\tau_N<\tau_{N-1}<\cdots<\tau_2<\tau_1\label{eq:old2-7}\end{equation}
without loss of generality. Then, we have $N$ real characteristic roots $\{\lambda_i\}_{i=1}^N$ structured in
\begin{equation}-\frac1{\tau_N}<\lambda_N<-\frac1{\tau_{N-1}}<\lambda_{N-1}<\cdots<-\frac1{\tau_2}<\lambda_2<-\frac1{\tau_1}<\lambda_1<0.\label{eq:interlacing}\end{equation}
The remaining two roots are either real negative roots or a complex conjugate pair of roots with negative real parts.
\end{theorem}
\begin{proof}Let $i=1,2,\ldots,N-1$ and consider the interval $(-1/\tau_{i+1},-1/\tau_i)$. Then, by taking \eqref{eq:old2-7} into account, we have
\begin{equation}P\left(-\frac1{\tau_i}\right)=(-1)^iG_i\left|\prod_{j\ne i}^N\left(1-\frac{\tau_j}{\tau_i}\right)\right|.\label{eq:old2-9}\end{equation}
Hence, by the intermediate value theorem, there exists a root $\lambda_{i+1}\in(-1/\tau_{i+1},-1/\tau_i)$.

Next, consider the interval $(-1/\tau_1,0)$. By \eqref{eq:old2-9} and \eqref{eq:poly}, we have $P_I(-1/\tau_1)<0$ and $P_I(0)=G_e>0$, respectively. Hence, by the intermediate value theorem, there exists a root $\lambda_1\in(-1/\tau_1,0)$. Thus, we have shown the existence of $N$ real characteristic roots $\{\lambda_i\}_{i=1}^N$ structured in \eqref{eq:interlacing}.

Finally, consider the remaining two roots, $\lambda_{N+1},\lambda_{N+2}$, and prove their real parts are negative. They must be either (i) complex conjugate pairs or (ii) purely real numbers. We first consider the case (i). Let $\lambda_{N+1,N+2}=p\pm iq$, where $p,q\in\mathbb R$ and $q\ne0$. Comparing the coefficients of $\lambda^{N+1}$ for $P_I(\lambda)$ and $I(\prod_{j=1}^N\tau_j)\prod_{i=1}^{N+2}(\lambda-\lambda_i)$, we have
\begin{equation}\sum_{j=1}^N\lambda_j+2p=-\sum_{j=1}^N\frac1{\tau_j}.\label{eq:old2-10}\end{equation}
Since $\lambda_j>-1/\tau_j$ for $1\le j\le N$, we have
\begin{equation}p=\frac12\sum_{j=1}^N\left(-\frac1{\tau_j}-\lambda_j\right)<0.\label{eq:old2-11}\end{equation}
As for the case (ii), just note that for $\lambda\ge0$,
\[{P_I}(\lambda)=(I\lambda^2+G_e)\prod_{j=1}^N(\tau_j\lambda+1)+\sum_{i=1}^N\left[G_i\tau_i\lambda\prod_{j=1,j\ne i}^N(\tau_j\lambda+1)\right]>0,\]
which means that any remaining real roots must be strictly negative.
\end{proof}
Introduce a notation $-1/\tau_0=0$ and consider each interval
$[-1/\tau_i,{-1/\tau_{i-1}}]$ for $i=1,2,\ldots,N$.
Note that each interval may have several roots of $P_I(\lambda)=0$.
A sufficient condition to avoid this case is that $P_I(\lambda)=0$ has a
non-real complex conjugate pair of roots. This is a useful
case for searching $\{\lambda_i\}_{i=1}^N$ using the bisection method,
which we will briefly discuss in Section 4. Concerning this, we have
the following theorem.
\begin{theorem}\label{thm:small}There exists a small positive $I_0$ such that if $0<I\le I_0$, then $P_I(\lambda)$ has exactly $N$ simple real roots $\{\lambda_i\}_{i=1}^N$ structured in \eqref{eq:interlacing}. The other two roots are simple complex conjugate roots with negative real parts.\end{theorem}
\begin{proof}We write $P_I(\lambda)$ in the form
\begin{equation}\begin{split}{P_I}(\lambda)={}&{G_e}\prod_{j=1}^N(\tau_j\lambda+1)+\sum_{i=1}^N\left[G_i\tau_i\lambda\prod_{j=1,j\ne i}^N(\tau_j\lambda+1)\right]+I\lambda^2\prod_{j=1}^N(\tau_j\lambda+1)\\={}&P^N(\lambda)+If(\lambda)\end{split}\label{eq:old2-12}\end{equation}
with
\begin{equation}\begin{cases}P^N(\lambda):=G_e\prod_{j=1}^N(\tau_j\lambda+1)+\sum_{i=1}^N\left[G_i\tau_i\lambda\prod_{j=1,j\ne i}^N(\tau_j\lambda+1)\right],\\f(\lambda):=\lambda^2\prod_{j=1}^N(\tau_j\lambda+1).\end{cases}\label{eq:old2-13}\end{equation}
We first show that $P^N(\lambda)$ has exactly $N$ real roots $\{\lambda_i\}_{i=1}^N$ in the interval $[-1/\tau_N,0]$. Taking \eqref{eq:old2-7} into account, we have
\begin{equation}P^N\left(-\frac1{\tau_i}\right)=(-1)^iG_i\left|\prod_{j\ne i}^N\left(1-\frac{\tau_j}{\tau_i}\right)\right|,\quad1\le i\le N.\label{eq:old2-14}\end{equation}
and $P^N(0)=G_e>0$, respectively. The intermediate value theorem gives a root in each of the $N$ disjoint intervals below. The polynomial $P^N$ has degree $N$ and leading coefficient $(G_e+\sum_iG_i)\prod_i\tau_i>0$. Hence these are all its roots, and each is simple. Thus
\begin{equation}-\frac1{\tau_N}<\lambda_N<-\frac1{\tau_{N-1}}<\lambda_{N-1}<\cdots<-\frac1{\tau_2}<\lambda_2<-\frac1{\tau_1}<\lambda_1<0.\label{eq:old2-15}\end{equation}
Set $\tau_0^{-1}=0$. By \eqref{eq:old2-15} and simplicity of the roots, we can take small $\delta>0$, $\epsilon>0$ such that
\begin{equation}{0<\delta<\frac12\min_{1\le j\le N}\{\lambda_j+\tau_j^{-1},-\tau_{j-1}^{-1}-\lambda_j\}},\label{eq:old2-16}\end{equation}
\begin{equation}|(P^N)'(\lambda)|\ge2\epsilon>0,\quad\lambda\in\cup_{j=1}^N[\lambda_j-\delta,\lambda_j+\delta],\label{eq:old2-17}\end{equation}
where we have abused the notation ``$'$'' to denote the derivative with respect to $\lambda$. The closed small intervals lie inside their respective interlacing intervals and are disjoint, also when $N=1$. Define the compact set
\[K=[-\tau_N^{-1},0]\setminus\bigcup_{j=1}^N(\lambda_j-\delta,\lambda_j+\delta).\]
Since $P^N$ has no zeros on $K$, let $m>0$ satisfy
\begin{equation}2m\le{\min_{\lambda\in K}}|P^N(\lambda)|,\label{eq:old2-18}\end{equation}
Put $M_0=\max_{[-\tau_N^{-1},0]}|f|$ and $M_1=\max_{[-\tau_N^{-1},0]}|f'|$. Choose
\[I_0=\min\{m/(1+M_0),\epsilon/(1+M_1)\}>0.\]
Then
\begin{equation}|If(\lambda)|\le m,\quad |If'(\lambda)|\le\epsilon,\quad\lambda\in[-1/\tau_N,0].\label{eq:old2-19}\end{equation}
hold for $0<I\le I_0$. It is easy to see that
\begin{equation}\begin{split}|{P_I}(\lambda)|&=|P^N(\lambda)+If(\lambda)|\\&\ge|P^N(\lambda)|-|If(\lambda)|\ge m,\quad{\lambda\in K}\end{split}\label{eq:old2-20}\end{equation}

and
\begin{equation}\begin{split}|{P_I^{\prime}}(\lambda)|&=|(P^N)'(\lambda)+If'(\lambda)|\\&\ge|(P^N)'(\lambda)|-|If'(\lambda)|\ge\epsilon,\quad\lambda\in\cup_{j=1}^N[\lambda_j-\delta,\lambda_j+\delta].\end{split}\label{eq:old2-21}\end{equation}
By \eqref{eq:old2-17}, $P^N$ is strictly monotone on each small interval. Its endpoint values have opposite signs. Their absolute values are at least $2m$ because both endpoints belong to $K$. By \eqref{eq:old2-19}, adding $If$ does not change these signs. Hence
\begin{equation}{P_I}(\lambda_j-\delta)\cdot {P_I}(\lambda_j+\delta)<0,\quad1\le j\le N.\label{eq:old2-22}\end{equation}
From \eqref{eq:old2-20}, \eqref{eq:old2-21} and \eqref{eq:old2-22}, there exists a single root of $P_I(\lambda)=0$ in $(\lambda_j-\delta,\lambda_j+\delta)$ for each $1\le j\le N$. These roots are simple by \eqref{eq:old2-21}, and \eqref{eq:old2-20} excludes all other roots in $[-\tau_N^{-1},0]$. Outside this interval, \eqref{eq:poly} gives
\[\frac{{P_I}(\lambda)}{\prod_i(1+\tau_i\lambda)}=I\lambda^2+G_e+\sum_i\frac{G_i\lambda}{\lambda+1/\tau_i}>0\quad\text{if }\lambda<-1/\tau_N\text{ or }\lambda\ge0.\]
Every ratio is nonnegative in these ranges and $G_e>0$. The poles are not roots by \eqref{eq:old2-9}. Thus $P_I$ has exactly $N$ real roots. Since it is a real polynomial of degree $N+2$, its remaining two roots form a nonreal conjugate pair. Both are simple, since higher multiplicity would require at least four remaining roots. Their real parts are negative by Theorem 2.1. Thus, we have completed the proof.
\end{proof}

\section{\texorpdfstring{Transfer function and recovery of characteristic roots}{Transfer function and recovery of characteristic roots}}
Consider the solution $\theta(t)$ of \eqref{eq:motion} with \eqref{eq:prony}, \eqref{eq:torque}, \eqref{eq:initial}. In this section, we first give an explicit form $\Theta(\lambda)=\mathcal L(\theta)(\lambda)$ of the Laplace transform of $\theta(t)$ including its analytic continuation. The Laplace transform of the prescribed torque $T(t)$ in \eqref{eq:torque} is
\begin{equation}{\widehat T(\lambda):=}\mathcal L\{T(t)\}(\lambda)=T_0\frac{\lambda\sin\phi+\omega_0\cos\phi}{\lambda^2+\omega_0^2}\label{eq:torque-laplace}\end{equation}
and
\begin{equation}\mathcal L\left(\exp\left(-\frac t{\tau_i}\right)\right)(\lambda)=\frac{{\tau_i}}{\tau_i\lambda+1}.\label{eq:exponential-laplace}\end{equation}
Then, using \eqref{eq:torque-laplace} and \eqref{eq:exponential-laplace}, we have
\begin{equation}{\Theta(\lambda)=\frac{\widehat T(\lambda)}{I\lambda^2+G_e+\displaystyle\sum_{i=1}^N\frac{G_i\tau_i\lambda}{\tau_i\lambda+1}}.}\label{eq:theta-laplace}\end{equation}
By \eqref{eq:theta-laplace}, define the transfer function by
\begin{equation}H_I(\lambda):=\frac{\Theta(\lambda)}{\widehat T(\lambda)}=\left[I\lambda^2+G_e+\sum_{i=1}^N\frac{G_i\tau_i\lambda}{1+\tau_i\lambda}\right]^{-1},\label{eq:transfer}\end{equation}
which is independent of the input $T(t)$ and its parameters $T_0$, $\omega_0$ and $\phi$.
With $D(\lambda)=\prod_{i=1}^N(1+\tau_i\lambda)$, \eqref{eq:poly} gives
\begin{equation}P_I(\lambda)=\left[I\lambda^2+G_e+\sum_{i=1}^N\frac{G_i\tau_i\lambda}{1+\tau_i\lambda}\right]D(\lambda).\label{eq:PI-D}\end{equation}
Together with \eqref{eq:transfer}, this gives $H_I(\lambda)=D(\lambda)/P_I(\lambda)$. Since $P_I(-a_i)\ne0$ for $a_i=1/\tau_i$ by \eqref{eq:old2-9}, $D(\lambda)$ and $P_I(\lambda)$ have no common roots. Hence the poles of $H_I(\lambda)$ are precisely the characteristic roots of the augmented system \eqref{eq:augmented}.

Next, we explain Pad\'e's method even for the case that there is an error in the measured angular displacement. This method recovers the poles of transfer function exactly and approximately corresponding to non-existence and existence of errors in measured angular displacements, respectively. Once again, we note here that the poles are the characteristic roots of $P_I$. For an observation time $T_{\mathrm{obs}}>0$, let the measured angular displacement be
\begin{equation}
\widetilde\theta(t)=\theta(t)+e(t),\qquad 0\le t\le T_{\mathrm{obs}},\qquad
\|e\|_{L^\infty(0,T_{\mathrm{obs}})}\le\varepsilon\|\theta\|_{L^\infty(0,T_{\mathrm{obs}})}.
\label{eq:measured-angle}
\end{equation}
Here $e(t)$ is the measurement error and $\varepsilon\ge0$ is its relative bound. To recover the relaxation function $G(t)$ from $\widetilde\theta(t)$, we first estimate the poles of the transfer function $H_I(\lambda)$ in \eqref{eq:transfer}, whose numerator and denominator have degrees $N$ and $N+2$, respectively.
For the prescribed frequency $\omega_0$ and phase $\phi$, choose an expansion point $\lambda_0>0$ such that $\widehat T(\lambda_0)\ne0$, equivalently, $\lambda_0\sin\phi+\omega_0\cos\phi\ne0$. Fix an expansion scale $h>0$, set $z=(\lambda-\lambda_0)/h$, and define $\mathcal H_I(z):=H_I(\lambda_0+hz)$.
It is known that Pad\'e's method \cite{pade} can be used to detect the poles of $H_I(\lambda)$ in the following steps.
\begin{enumerate}\renewcommand\labelenumi{(\roman{enumi})}
\item Compute the power series of $\mathcal H_I(z)$ at $z=0$ up to order $2N+2$. Denote this finite series by $M(z)=\sum_{k=0}^{2N+2}c_kz^k$, which approximates $\mathcal H_I(z)=H_I(\lambda_0+hz)$ near $z=0$.
Use the measured angular displacement $\widetilde\theta(t)$ in \eqref{eq:measured-angle} and the prescribed torque $T(t)$ in \eqref{eq:torque}. Estimate the coefficients $b_k$ of Maclaurin's expansion of $\Theta(\lambda_0+hz)$ at $z=0$ approximately and compute the coefficients $c_k$ of $\mathcal H_I(z)$ recursively from $\widehat T H_I=\Theta$:
\begin{equation}
\begin{aligned}
b_k&=\frac{(-h)^k}{k!}\int_0^{T_{\mathrm{obs}}}t^ke^{-\lambda_0t}\widetilde\theta(t)\,dt,\\
c_k&=\frac{(\lambda_0^2+\omega_0^2)b_k+2\lambda_0h b_{k-1}+h^2b_{k-2}-T_0h\sin\phi\,c_{k-1}}{T_0(\lambda_0\sin\phi+\omega_0\cos\phi)}.
\end{aligned}\label{eq:angular-coefficients}
\end{equation}
Here $0\le k\le2N+2$, and coefficients with negative subscripts are zero. The computed coefficients $c_k$ include quadrature, truncation of the observation interval and measurement errors.
\item Set up Pad\'e's approximant $A(z)/Q_{\mathrm P}(z)$ of $\mathcal H_I(z)$, where
\[A(z)=\sum_{j=0}^N\alpha_jz^j,\qquad Q_{\mathrm P}(z)=\sum_{j=0}^{N+2}\beta_jz^j,\]
with constants $\alpha_0,\ldots,\alpha_N$ and $\beta_j$'s.
\item Determine these constants to satisfy
\begin{equation}
M(z)Q_{\mathrm P}(z)-A(z)=O(z^{2N+3}),\label{eq:single-pade}
\end{equation}
Since $\deg A\le N$, comparing coefficients of $z^{N+1},\ldots,z^{2N+2}$ in \eqref{eq:single-pade} gives the equations for the denominator coefficients $\beta_j$:
\begin{equation}\sum_{j=0}^{N+2}c_{k-j}\beta_j=0,\qquad N+1\le k\le 2N+2,\label{eq:pade-denominator}\end{equation}
where $c_k=0$ for $k<0$ and $\beta_0=1$.
\item Compute the roots $z_j$ of $Q_{\mathrm P}(z)$ and obtain the estimated characteristic roots $\widetilde\lambda_j=\lambda_0+hz_j$.
\end{enumerate}

\noindent\textit{Remark.} Pad\'e's SVD algorithm of Gonnet et al.~\cite{gonnet} is applied to the approximate coefficients $c_k$ computed from noisy angular displacement measurements $\widetilde\theta(t)$ by \eqref{eq:angular-coefficients}. It uses the singular values of the coefficient matrix of \eqref{eq:pade-denominator} to determine its numerical rank and computes the denominator coefficients $\beta_j$ from a null vector.

\section{\texorpdfstring{Inversion algorithm and numerical experiments}{Inversion algorithm and numerical experiments}}
The reconstruction has two steps. First, recover the characteristic roots $\lambda_j^I,\lambda_j^{I^{\prime}}$ from the angular displacements $\theta_I(t),\theta_{I^{\prime}}(t)$ using Pad\'e's approximant as described in Section~3. Second, use these roots $\lambda_j^I,\lambda_j^{I^{\prime}}$ to recover the relaxation function $G(t)$ by the mentioned spectral approach. The numerical experiment applies both steps consecutively. We note here that the robustness of the second step against noise is reported in \cite{chen2}.

\subsection{\texorpdfstring{Reconstruction of relaxation parameters from characteristic roots}{Reconstruction of relaxation parameters from characteristic roots}}\label{sec:numerical-experiments}
We apply the spectral approach to reconstruct the relaxation parameters. To begin with, for each fixed inertia $I$, let $\lambda_j^I$, $j=1,\ldots,N+2$, be the roots of $P_I$. This finite set plays the role of one cluster in \mbox{\cite{chen1,chen2}}. A complete set of characteristic roots means all roots in this cluster. The second step uses the two sets of roots recovered in the first step as input. For approximate roots, the numerical implementation and its checks are given below.
Recall the well-known fact that for a polynomial having the coefficient equal 1 for the term with the highest degree, a monic polynomial, knowing all of its roots is equivalent to knowing the polynomial. Needless to say that a similar fact is also true for the case when the coefficient 1 is replaced by a known nonzero constant.

Let $I\ne I^{\prime}$ be two known positive moments of inertia for the same material. Assume that $N$ is known, $G_e,G_i>0$, and the relaxation times $\tau_i>0$ are distinct.

The reconstruction proceeds in four steps.

\begin{enumerate}
\item Form the monic characteristic polynomials $\chi_I(\lambda)$ and $\chi_{I^{\prime}}(\lambda)$ from the characteristic roots $\lambda_j^I$ and $\lambda_j^{I^{\prime}}$, respectively:
\[
\chi_I(\lambda)=\prod_{j=1}^{N+2}(\lambda-\lambda_j^I),\qquad
\chi_{I^{\prime}}(\lambda)=\prod_{j=1}^{N+2}(\lambda-\lambda_j^{I^{\prime}}).
\]
By \eqref{eq:det-poly}, $P_I(\lambda)=I(\prod_{j=1}^N\tau_j)\chi_I(\lambda)$ and $P_{I^{\prime}}(\lambda)=I^{\prime}(\prod_{j=1}^N\tau_j)\chi_{I^{\prime}}(\lambda)$.
\item Construct $R(\lambda)$ from $\chi_I(\lambda)$ and $\chi_{I^{\prime}}(\lambda)$. Equations \eqref{eq:poly} and \eqref{eq:det-poly} give
\begin{equation}
R(\lambda)=\frac{I\chi_I(\lambda)-I^{\prime}\chi_{I^{\prime}}(\lambda)}{(I-I^{\prime})\lambda^2}
=\prod_{j=1}^N(\lambda+1/\tau_j).
\label{eq:recover}\end{equation}
Here $R(0)=\prod_{j=1}^N\tau_j^{-1}>0$.
\item Form the polynomial $Q(\lambda,I,I^{\prime})=R(\lambda)/R(0)$. By \eqref{eq:poly} and the relations above,
\begin{equation}
Q(\lambda,I,I^{\prime}):=\frac{R(\lambda)}{R(0)}=\frac{P_I(\lambda)-P_{I^{\prime}}(\lambda)}{(I-I^{\prime})\lambda^2}=D(\lambda).
\label{eq:original-Q}
\end{equation}
Since $D(\lambda)=\prod_{i=1}^N(1+\tau_i\lambda)$, the roots of $Q$ in \eqref{eq:original-Q} are $-a_i=-1/\tau_i$. Find the roots $-a_N<\cdots<-a_1<0$ and recover the relaxation times $\tau_i=1/a_i$.

Under Theorem~\ref{thm:small}, the real characteristic roots $\lambda_N^I<\cdots<\lambda_1^I$ in \eqref{eq:interlacing} determine intervals for the bisection \mbox{method \cite{burden}}. For $j=2,\ldots,N$, the interval $(\lambda_j^I,\lambda_{j-1}^I)$ contains exactly one root $-a_{j-1}$ of $Q$, and
\[
Q(\lambda_j^I,I,I^{\prime})Q(\lambda_{j-1}^I,I,I^{\prime})<0.
\]
For the root $-a_N=-1/\tau_N$, write $R(\lambda)=\lambda^N+q_{N-1}\lambda^{N-1}+\cdots+q_0$ and define the positive number $L=2+\max_{0\le k<N}|q_k|$. Use the interval $(-L,\lambda_N^I)$ to find the root $-a_N$ of $Q(\lambda,I,I^{\prime})$ by the bisection method. The Cauchy root bound theorem for $R$ gives $a_N<L$, hence $-L<-a_N$, while \eqref{eq:interlacing} gives $-a_N<\lambda_N^I$. No other root of $Q$ lies in this interval, so $Q(-L,I,I^{\prime})Q(\lambda_N^I,I,I^{\prime})<0$. This also covers $N=1$.
\item Recover the moduli $G_i$ and $G_e$ using the following relations obtained from \eqref{eq:poly}:
\begin{equation}
G_i=-\frac{I\chi_I(-a_i)}{R(0)\prod_{j\ne i}(1-\tau_j/\tau_i)},\qquad G_e=\frac{I\chi_I(0)}{R(0)}.
\label{eq:original-moduli}
\end{equation}
Here $R(0)=\prod_{i=1}^N a_i$ is constructed from the recovered rates $a_i$.
Substitute $\tau_i,G_i,G_e$ into \eqref{eq:prony} to obtain $G(t)$.
\end{enumerate}

\subsection{\texorpdfstring{Reconstruction of relaxation parameters from angular displacements}{Reconstruction of relaxation parameters from angular displacements}}\label{sec:joint-recovery}

We recover the characteristic roots $\lambda_j^I,\lambda_j^{I^{\prime}}$ from synthetic angular displacements $\theta_I(t),\theta_{I^{\prime}}(t)$. The relaxation rates are $a_i=1/\tau_i$ for $i=1,2$. Take $N=2$, $G_e=2$, $(G_1,G_2)=(1,2)$, $(a_1,a_2)=(0.2,2)$, $I=0.05$ and $I^{\prime}=0.2$. The rates $a_1<a_2$ correspond to $\tau_1=5>\tau_2=0.5$, following the ordering of the relaxation times as in Section~2. The choice $G_1<G_2$ gives a larger contribution to the faster relaxing term in \eqref{eq:prony}. This choice is specific to the example and is not required by the model. In \eqref{eq:torque}, take $T_0=1$, $\phi=0$ and $\omega_0\in\{1,100\}$, so the input is $T(t)=\sin(\omega_0t)$. The angular displacement $\theta_k(t)$ is sampled at $t_m=0.005m$, $m=0,\ldots,40000$, for $k\in\{I,I^{\prime}\}$. The noisy samples $\widetilde\theta_k(t_m)$ are
\[\widetilde\theta_k(t_m)=\theta_k(t_m)+\varepsilon U_{k,m}\max_{0\leq l\leq 40000}|\theta_k(t_l)|,\qquad U_{k,m}\sim\mathrm{Uniform}[-1,1],\]
where $U_{k,m}$ are independent across sample times and inertia settings.

For each set of angular displacement samples $\widetilde\theta_I(t_m)$ or $\widetilde\theta_{I^{\prime}}(t_m)$, trapezoidal quadrature in \eqref{eq:angular-coefficients} gives the coefficients $b_k$. The same equation then gives $c_k$, with $\lambda_0=h=1$. The SVD algorithm in \cite{gonnet} computes the Pad\'e denominator $Q_{\mathrm P}(z)$ and its roots $z_j$, using relative tolerance $10^{-14}$. The eigenvalues $\lambda_j^I,\lambda_j^{I^{\prime}}$ of the augmented matrices $\mathbf A_I,\mathbf A_{I^{\prime}}$ provide the reference roots. There is one noise free trial and 100 trials at each nonzero noise level. For each inertia $k\in\{I,I^{\prime}\}$, let $\widetilde\lambda_j^k$ denote the recovered roots and $\lambda_j^k$ the reference roots.

\begin{figure}[!htbp]
\centering
\includegraphics[width=0.9\textwidth]{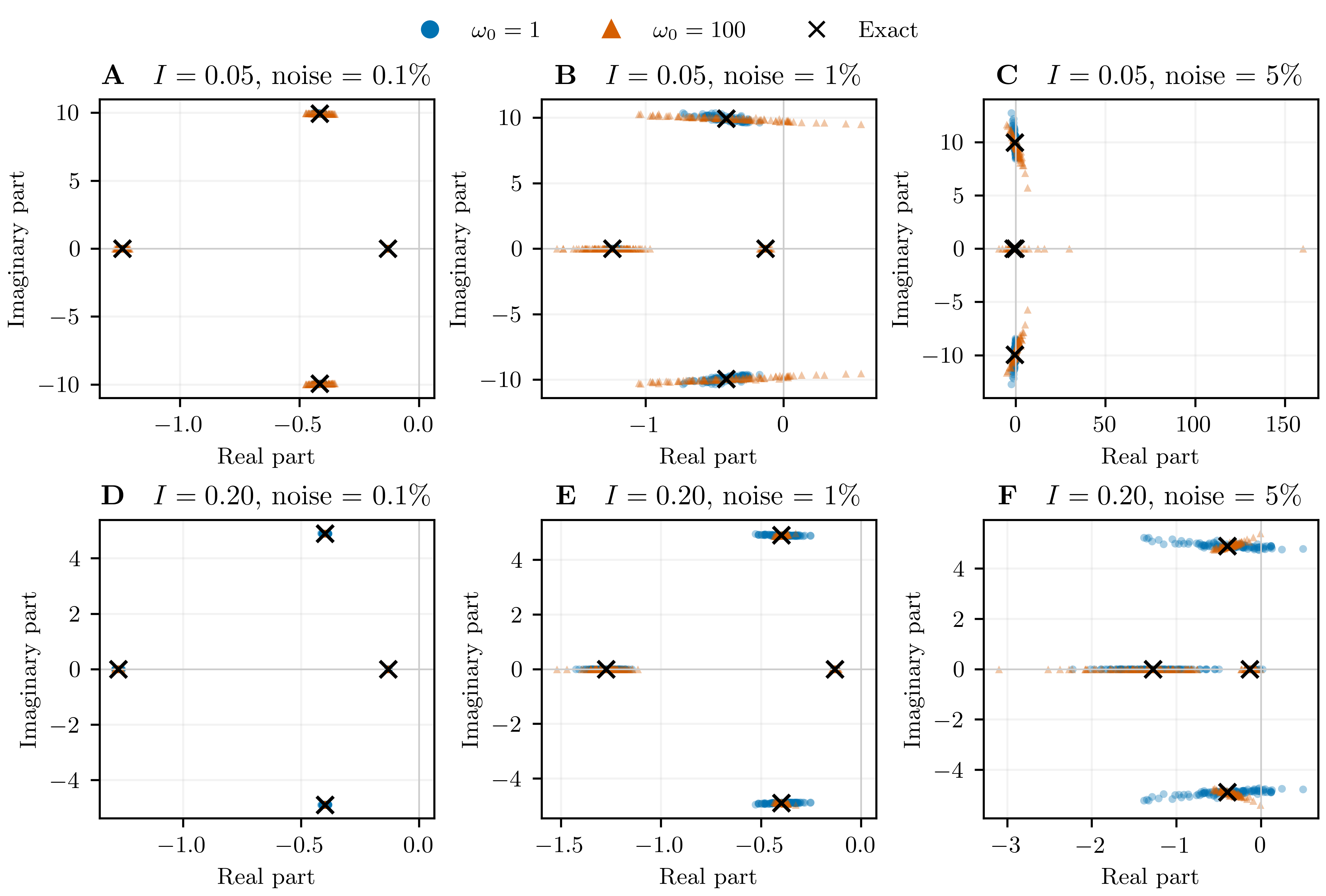}
\caption{Characteristic roots $\widetilde\lambda_j^I,\widetilde\lambda_j^{I^{\prime}}$ recovered from noisy angular displacements. The crosses show the exact roots. Blue circles and orange triangles show the recovered roots for $\omega_0=1$ and $\omega_0=100$, respectively, from all 100 trials at each noise level.}
\label{fig:recovered-roots}
\end{figure}

Figure~\ref{fig:recovered-roots} shows the recovered characteristic roots $\widetilde\lambda_j^I,\widetilde\lambda_j^{I^{\prime}}$. For each trial, the maximum relative root error $E_\lambda$ is
\[
E_\lambda=\max_{k\in\{I,I^{\prime}\}}\min_{\pi\in\mathfrak S_{N+2}}\max_{1\le j\le N+2}\frac{|\widetilde\lambda_j^k-\lambda_{\pi(j)}^k|}{|\lambda_{\pi(j)}^k|},
\]
where $\mathfrak S_{N+2}$ is the set of permutations of $\{1,\ldots,N+2\}$.

For $k\in\{I,I^{\prime}\}$, the recovered characteristic roots $\widetilde\lambda_j^k$ determine the polynomials $\widetilde\chi_k(\lambda)=\prod_{j=1}^{N+2}(\lambda-\widetilde\lambda_j^k)$. The bisection method is applied to $I\widetilde\chi_I-I^{\prime}\widetilde\chi_{I^{\prime}}$ to obtain the rates $\widehat a_i$ and hence the relaxation times $\widehat\tau_i=1/\widehat a_i$. Unlike the exact difference in \eqref{eq:recover}, this polynomial need not be divisible by $\lambda^2$. The moduli $\widehat G_i$ and $\widehat G_e$ are then recovered using \eqref{eq:original-moduli}. Missing brackets, failed root-count checks, nonpositive parameters and nonfinite values are classified as invalid.

The reconstruction errors for the parameter vector $\mathbf p=(G_e,G_1,G_2,a_1,a_2)$ and the relaxation function $G(t)$ are
\[E_p=\max_k|\widehat p_k/p_k-1|,\qquad E_G=\frac{\|(\widehat G(t_j)-G(t_j))_{j=1}^{400}\|_2}{\|(G(t_j))_{j=1}^{400}\|_2},\]
where $t_j$ are logarithmically spaced on $[0.01,100]$. The median $E_\lambda$ is calculated from all 100 trials. The medians $E_p$ and $E_G$ are calculated from valid reconstructions only.

\begin{figure}[H]
\centering
\includegraphics[width=0.85\textwidth]{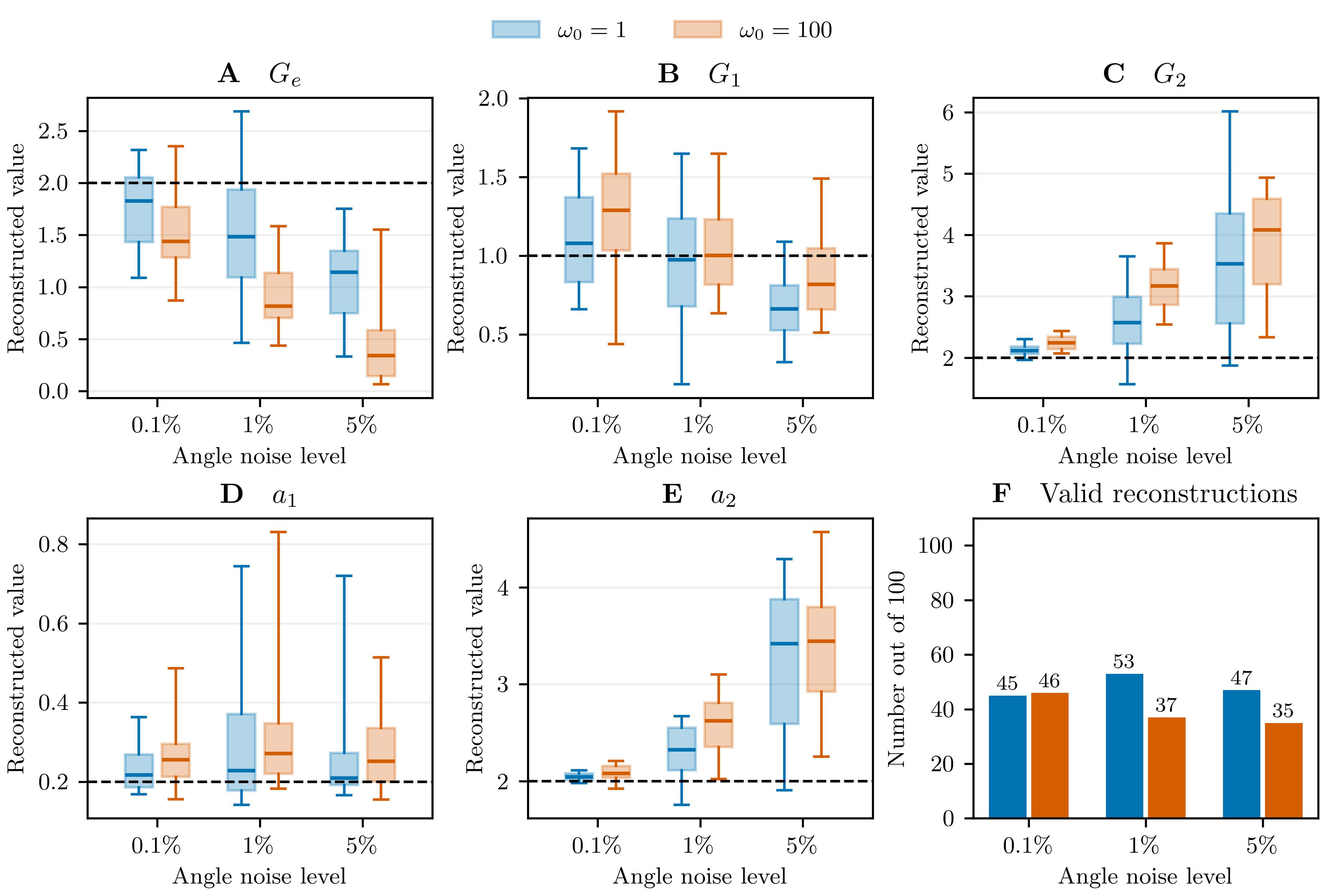}
\caption{Relaxation parameters $\widehat G_e,\widehat G_i,\widehat a_i$ reconstructed from angular displacements. Blue and orange indicate $\omega_0=1$ and $\omega_0=100$, respectively. The boxplots use valid reconstructions only. Boxes show the first and third quartiles, horizontal lines show medians, whiskers show the 5th and 95th percentiles, and dashed lines show the exact values. The numbers above the plots give valid counts out of 100.}
\label{fig:joint-parameters}
\end{figure}

Figure~\ref{fig:joint-parameters} shows the reconstructed parameters $\widehat G_e,\widehat G_i,\widehat a_i$ of the relaxation function $G(t)$. For the noise free synthetic angular displacements, the reconstructed parameter vector $\widehat{\mathbf p}=(\widehat G_e,\widehat G_1,\widehat G_2,\widehat a_1,\widehat a_2)$ is $(2.000,1.000,2.000,0.200,2.000)$ for both $\omega_0=1$ and $\omega_0=100$, rounded to three decimal places. The exact vector $\mathbf p=(G_e,G_1,G_2,a_1,a_2)$ is $(2,1,2,0.2,2)$.

\begin{table}[H]
\centering

\caption{Median reconstruction errors and numbers of valid reconstructions out of 100 trials.}
\label{tab:reconstruction-errors}
\begin{tabular}{ccrrrr}
\hline
$\omega_0$ & $\varepsilon$ & $E_\lambda$ & Valid reconstructions & $E_p$ & $E_G$ \\
\hline
1 & 0.1\% & 0.00406 & 45/100 (45\%) & 0.269 & 0.0617 \\
1 & 1\% & 0.03676 & 53/100 (53\%) & 0.51 & 0.151 \\
1 & 5\% & 0.19307 & 47/100 (47\%) & 0.778 & 0.276 \\
\hline
100 & 0.1\% & 0.01116 & 46/100 (46\%) & 0.384 & 0.113 \\
100 & 1\% & 0.11459 & 37/100 (37\%) & 0.722 & 0.262 \\
100 & 5\% & 0.68499 & 35/100 (35\%) & 1.114 & 0.423 \\
\hline
\end{tabular}
\end{table}

Table~\ref{tab:reconstruction-errors} and Figures~\ref{fig:recovered-roots}--\ref{fig:joint-parameters} give the following observations. The parameter errors below are medians of $|\widehat p_k/p_k-1|$ over valid trials, whose counts are reported in Table~\ref{tab:reconstruction-errors}.
\begin{itemize}
\item Without noise, $\widehat{\mathbf p}$ agrees with $\mathbf p$ to three decimal places for both frequencies $\omega_0$. For noisy data, $\omega_0=1$ gives smaller median errors $E_\lambda$, $E_p$ and $E_G$ than $\omega_0=100$ in the present experiment. Although $H_I(\lambda)$ is independent of $T(t)$, its estimation from finite noisy angular displacements depends on the input.
\item At $0.1\%$ noise, the fast relaxation parameters $a_2$ and $G_2$ are recovered more accurately than $a_1$, $G_1$ and $G_e$. For $\omega_0=1$, their median relative errors are $0.022$ and $0.056$, compared with $0.157$, $0.269$ and $0.165$, respectively. The larger modulus $G_2$ gives the fast term a greater initial contribution to $G(t)$. The exponential weight in \eqref{eq:angular-coefficients} also suppresses later observations that help distinguish slow relaxation. These features may contribute to the difference in accuracy.
\item The median function error $E_G$ is smaller than the median parameter error $E_p$ at every tested noise level $\varepsilon$. For example, they are $0.151$ and $0.510$ at $\omega_0=1$ and $1\%$ noise. The estimates $\widehat G_2$ and $\widehat a_2$ tend to increase together as $\varepsilon$ grows. Their changes can partly compensate in the integrated contribution $\int_0^\infty G_2e^{-a_2t}\,dt=G_2/a_2$. This provides one possible explanation for the better recovery of $G(t)$, although $E_G$ and $E_p$ measure different quantities.
\item The parameters also enter different reconstruction formulas. The rates $\widehat a_i$ are obtained by finding roots, whereas \eqref{eq:original-moduli} recovers the moduli $\widehat G_i$ and $\widehat G_e$ by polynomial evaluation. In particular, $G_e=I\chi_I(0)/R(0)$ with $R(0)=\prod_i a_i$ combines information from the characteristic roots and the recovered rates. This dependence helps explain the different parameter distributions in Figure~\ref{fig:joint-parameters}. Comparisons across noise levels concern the corresponding valid trials.\end{itemize}

\FloatBarrier

\section{\texorpdfstring{Conclusion and future works}{Conclusion and future works}}

This paper develops a method to recover a relaxation function $G(t)$ represented by a Prony's series with a positive constant term. For known order $N$ and distinct relaxation times $\tau_i$, two complete sets of characteristic roots $\lambda_j^I$ and $\lambda_j^{I^{\prime}}$ uniquely determine $G(t)$ through \eqref{eq:recover}--\eqref{eq:original-moduli}. Pad\'e's approximation is used to estimate the poles of the transfer function, which are nothing but the characteristic roots from angular displacements.

Numerical experiments apply both steps to synthetic angular displacements. They confirm recovery of the relaxation parameters from noise free data. For valid reconstructions from noisy data, Table~\ref{tab:reconstruction-errors} shows that the median function errors $E_G$ are smaller than the parameter errors $E_p$.

Future work will address the following.
\begin{itemize}
\item Study the stability of Pad\'e's approximation for recovering characteristic roots from noisy angular displacement measurements and regularization of the parameter reconstruction.
\item Examine inertia calibration, synchronized torque and angular displacement measurements, and feedback effects in experiments.
\end{itemize}

\section*{Acknowledgments}
The authors thank Dr.\ Shuli Li for providing the bisection code used as a reference for the numerical implementation. The first author was supported by the National Natural Science Foundation of China (Grant No. 12241103). The third author was supported by the Ministry of Science and Technology of Taiwan (Grant No. NSTC 111-2115-M-006-013-MY3). The fourth author was supported by JSPS KAKENHI (Grant No. JP25K07076).


\begin{thebibliography}{99}
\bibitem{rheocompass} Anton Paar GmbH, RheoCompass Software: Software Manual, document C01IB003EN-F, June 2018, \url{https://12idb.xray.aps.anl.gov/files/Rheo_MCR702e_Software.pdf}.
\bibitem{pade} G. Baker, Jr. and P. Graves-Morris, Pad\'e Approximants, 2nd ed., Cambridge University Press, 1996.
\bibitem{burden} R. Burden and J. Faires, Numerical Analysis, Cengage Learning, Boston, 2010.
\bibitem{chen1} S. Chen, M. de Hoop, Y. Deng, C.-L. Lin, G. Nakamura, Clustered eigenvalue problem for glassy state relaxation and its inverse problem, \emph{SIAM J. Appl. Math.}, \textbf{86} (2026), 1058--1075.
\bibitem{chen2} S. Chen, M. de Hoop, Y. Deng, C.-L. Lin, G. Nakamura, Inverse spectral problem for glassy state relaxation approximated by Prony series, \emph{Appl. Math. Lett.}, \textbf{178} (2026), Article 109904.
\bibitem{ewoldt} R. Ewoldt and G. McKinley, Creep ringing in rheometry or how to deal with oft-discarded data in step stress tests!, \emph{Rheol. Bull.}, \textbf{76}(1) (2007).
\bibitem{geri} M. Geri, B. Keshavarz, T. Divoux, C. Clasen, D. Curtis and G. McKinley, Time-Resolved Mechanical Spectroscopy of Soft Materials via Optimally Windowed Chirps, \emph{Phys. Rev. X}, \textbf{8} (2018), Article 041042, doi:10.1103/PhysRevX.8.041042.
\bibitem{gonnet} P. Gonnet, S. G\"uttel and L. Trefethen, Robust Pad\'e approximation via SVD, \emph{SIAM Rev.}, \textbf{55} (2013), 101--117.
\bibitem{hamley} I. Hamley, Introduction to Soft Matter, John Wiley, New Jersey, 2007.
\bibitem{honerkamp} J. Honerkamp and J. Weese, A nonlinear regularization method for the calculation of relaxation spectra, \emph{Rheol. Acta}, \textbf{32} (1993), 65--73, doi:10.1007/BF00396678.
\bibitem{kammler} D. Kammler, Prony's method for completely monotonic functions, \emph{J. Math. Anal. Appl.}, \textbf{57} (1977), 560--570.
\bibitem{lauger2010} J. L\"auger and H. Stettin, Differences between stress and strain control in the non-linear behavior of complex fluids, \emph{Rheol. Acta}, \textbf{49} (2010), 909--930, doi:10.1007/s00397-010-0450-0.
\bibitem{lauger2016} J. L\"auger and H. Stettin, Effects of instrument and fluid inertia in oscillatory shear in rotational rheometers, \emph{J. Rheol.}, \textbf{60} (2016), 393--406, doi:10.1122/1.4944512.
\bibitem{luo} R. Luo, H. Lv, H. Liu, Development of Prony series models based on continuous relaxation spectrums for relaxation moduli determined using creep tests, \emph{Constr. Build. Mater.}, \textbf{168} (2018) 758--770.
\bibitem{mauro} J. Mauro and Y. Mauro, On the Prony series representation of stretched exponential relaxation, \emph{Physica A}, \textbf{506} (2018), 75--87, doi:10.1016/j.physa.2018.04.047.
\bibitem{ogunkeye} A. Ogunkeye, R. Hudson and D. Curtis, The effect of instrument inertia on the initiation of oscillatory flow in stress controlled rheometry, \emph{J. Rheol.}, \textbf{67} (2023), 1175--1187, doi:10.1122/8.0000665.
\bibitem{potts} D. Potts and M. Tasche, Parameter estimation for exponential sums by approximate Prony method, \emph{Signal Process.}, \textbf{90} (2010), 1631--1642, doi:10.1016/j.sigpro.2009.11.012.
\bibitem{training} TA Instruments, Rheology Theory and
Applications training slides, hosted by Clarkson University,
\url{https://people.clarkson.edu/~skrishna/DHR_Rheology_Theory.pdf}.
\end{thebibliography}
\end{document}